\documentclass[3p,times,procedia]{elsarticle}
\usepackage{nupha_ecrc}

\volume{00}
\firstpage{1}
\journalname{Nuclear Physics A}
\runauth{Lipei Du et al.}
\jid{nupha}
\jnltitlelogo{Nuclear Physics A}

\usepackage{amssymb}
\usepackage{lineno}
\usepackage{multicol}
\usepackage{graphicx}
\usepackage{bm}

\usepackage{amsmath,latexsym}
\usepackage{subfigure}
\usepackage{slashed}
\usepackage{multirow,array}
\usepackage{mathtools}
\usepackage{mathrsfs}
\usepackage[colorlinks=false,linktocpage=true]{hyperref}
\usepackage{hyperref}
\usepackage[utf8]{inputenc}
\usepackage{lipsum}
\usepackage[rightcaption]{sidecap}

\usepackage[usenames]{color}
\definecolor{darkblue}{RGB}{0,0,196}
\definecolor{darkred}{RGB}{196,0,0}

\biboptions{comma,sort&compress}

\usepackage[figuresright]{rotating}

\newcommand{\br}{\bm{r}}
\newcommand{\bp}{\bm{p}}

\newcommand{\bu}{\bm{u}}

\newcommand{\V}{n}
\newcommand{\n}{{\rho_B}}
\newcommand{\ed}{{e}}

\newcommand{\peq}{{p}_{0}}

\begin{document}

\begin{frontmatter}

%% Title, authors and addresses

\dochead{XXVIIth International Conference on Ultrarelativistic Nucleus-Nucleus Collisions\\ (Quark Matter 2018)}

\title{Hybrid model with dynamical sources for heavy-ion collisions at BES energies\tnoteref{tn1}}
\tnotetext[tn1]{Supported by DOE (award no.\,DE-SC0004286 and BEST Collaboration) and NSF (JETSCAPE Collaboration, ACI-1550223).}

%% use optional labels to link authors explicitly to addresses:
\author[label1]{Lipei Du\footnote{Presenter. E-mail: du.458@osu.edu.}}
\author[label1,label2,emmi]{Ulrich Heinz}
\author[label1]{Gojko Vujanovic}
\address[label1]{Department of Physics, The Ohio State University, Columbus, OH 43210-1117, USA}
\address[label2]{Theoretical Physics Department, CERN, CH-1211 Geneve 23, Switzerland}
\address[emmi]{ExtreMe Matter Institute (EMMI), GSI Helmholtzzentrum f\"ur Schwerionenforschung, 
                          Planckstrasse 1, D-64291 Darmstadt, Germany\\[-3ex]}

\begin{abstract}
We develop a (3+1)-dimensional hybrid evolution model for heavy-ion collisions with dynamical sources for the energy-momentum tensor and baryon current. During an initial pre-equilibrium stage based on UrQMD, the four-momenta and baryon numbers carried by secondary particles created within UrQMD are fed continuously, after a short thermalization time, into a (3+1)-dimensional viscous hydrodynamic evolution module including baryon transport. The sensitivity of the initial conditions to model parameters and the effect of baryon diffusion on the hydrodynamic evolution are studied.
\end{abstract}


\end{frontmatter}

%%%%%%%%%%%%%%%%%%%%%%%%%%%%%%%%%%%%%%%%%%%%%%%%%%%%%%%%%%%%%%%%%%%%%%%%%%
%% main text

%%%%%%%%%%%%%%%%%%%%%%%%%%%%%%%%%%%%%%%%%%%%
\section{Introduction}
\label{sec1}
%%%%%%%%%%%%%%%%%%%%%%%%%%%%%%%%%%%%%%%%%%%%

At Relativistic Heavy Ion Collider (RHIC) Beam Energy Scan (BES) energies, the dynamics of the pre-equilibrium stage and the effects resulting from a nonzero net baryon current become critical components of the dynamical evolution of the collision fireball \cite{Shen:2017ruz}. Recently hybrid models of heavy-ion collisions, consisting of multiple stages describing different physics, have received intensive attention. In many approaches (see Table I in \cite{Oliinychenko:2015lva}), a hydrodynamic stage describing the evolution of quark-gluon plasma is initialized with output from some pre-equilibrium evolution model on a surface of constant (proper) time. Dynamical initialization models in which the pre-equilibrium matter is converted to fluid gradually while the colliding nuclei are passing through each other were proposed in \cite{Shen:2017ruz, Okai:2017ofp, Shen:2017bsr, Akamatsu:2018olk}. In this work, we study the dynamical initialization of hydrodynamics from UrQMD \cite{Bass:1998ca}. Our approach has many similarities with \cite{Akamatsu:2018olk} (where JAM was used instead of UrQMD) but, different from \cite{Akamatsu:2018olk} and similar to \cite{Denicol:2018wdp}, it uses dissipative hydrodynamics, including evolution of the baryon diffusion current. We will here focus on differences between the initial conditions obtained from our approach and that of \cite{Shen:2017bsr,Denicol:2018wdp}, and on the dynamical effects of baryon number diffusion which were studied in \cite{Denicol:2018wdp} but not in \cite{Akamatsu:2018olk}.

%%%%%%%%%%%%%%%%%%%%%%%%%%%%%%%%%%%%%%%%%%%%
\section{Pre-equilibrium dynamics and dynamical sources}
\label{sec2}
%%%%%%%%%%%%%%%%%%%%%%%%%%%%%%%%%%%%%%%%%%%%

During the interpenetration stage of the two nuclei, we describe the medium created in the collision as a superposition of freshly produced, still un-thermalized particles and an approximately thermalized dissipative fluid. In the conservation laws for energy, momentum and net baryon number, hydrodynamic source currents describe the conversion of particles into fluid via thermalization:
\begin{equation}
\label{eq1}
   \partial_\mu T^{\mu\nu}_{\mathrm{fluid}} 
   = J_{\mathrm{source}}^\nu(x) \equiv -\partial_\mu T^{\mu\nu}_{\mathrm{particle}}(x) \;,\quad 
   \partial_\mu N^\mu_{\mathrm{fluid}} = \rho_{\mathrm{B}{\textrm{source}}}(x)  \equiv -\partial_\mu N^\mu_{\mathrm{particle}}(x)\;,
\end{equation}
where $\partial_\mu$ stands for the covariant derivative. The particle contributions on the right hand side are obtained from UrQMD \cite{Bass:1998ca}, a kinetic model based on hadronic degrees of freedom that describes the initial collision stage in terms of the decay of strings and resonances created in the primary collisions between nucleons as the colliding nuclei interpenetrate each other. With the exception of leading baryons carrying at least one of the incoming valence quarks, particles produced in these decays are not allowed to rescatter but assumed to become part of the fluid after free-streaming for a formation time $\tau_{\mathrm{form}}$ which encapsulates in a single, species-independent number both their formation and thermalization, in their own rest frame. Leading baryons are allowed to scatter multiple times if the secondary collision occurs within their formation time, until the nuclei have completely passed through each other; then they, too, become part of the fluid.

The energy-momentum tensor and net baryon current of the particles thus produced are given by \cite{Oliinychenko:2015lva}
\begin{equation}
\label{eq2}
   T^{\mu\nu}_{\mathrm{particle}}(t, \br) 
   = \sum_i \frac{p_i^\mu p_i^\nu}{p_i^0} K(\br{-}\br_i(t), \bp_i)\,\Theta_i\;, \quad
   N^\mu_{\textrm{particle}}(t, \br) 
   = \sum_i b_i \frac{p_i^\mu}{p_i^0} K(\br{-}\br_i(t), \bp_i)\,\Theta_i\;,
\end{equation}
where $p_i^0{\,=}\sqrt{m_i^2{\,+\,}\bp_i^2}$, $\br_i(t)=\br_{i0}+\bigl(\bp_i/p_i^0\bigr)(t{\,-\,}t_{i0})$ is the free-streaming trajectory of a particle produced at space-time point $x_{i0}^\mu=(t_{i0},\br_{i0})$. Here $K(\br{-}\br_i(t),\bp_i)$ and $\Theta_i{\,\equiv\,}\Theta(t_{\mathrm{form},i}{\,-\,}t_{\mathrm{rf}})$ are a spatial smearing kernel, assumed to be Gaussian, and a step-like temporal switching function in the rest frame (rf) of the particle, respectively:
\begin{equation}
   K(\br,\bp_i) = \frac{\gamma_i}{\left(2\pi\sigma^2\right)^{3/2}}
   \exp\left(-\frac{\br^2 + (\br\cdot\bu_i)^2}{2\sigma^2}\right),\qquad	
   \Theta(t_{\textrm{form}}{\,-\,}t)
   =\frac{1}{2}\left[\tanh\left(\frac{t_{\textrm{form}}-t}{\Delta \tau_{\textrm{th}}}\right)+1\right].
\end{equation}
%
%%%%%%%%%%%%%%%%%% Fig. 1 %%%%%%%%%%%%%%%%%%%%%%%%%%%%%
\begin{figure}[b]
\centering
\includegraphics[width=0.45\linewidth]{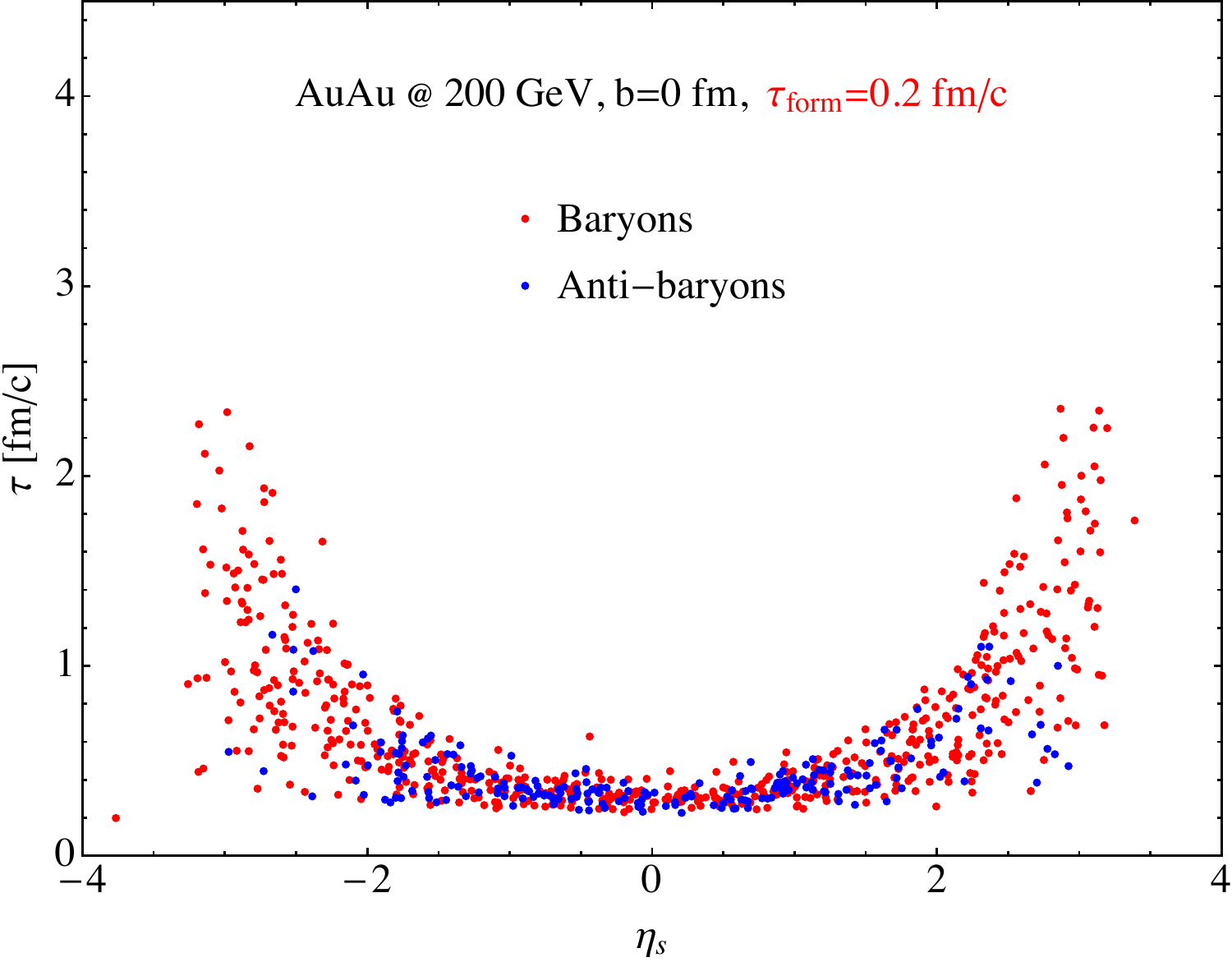}\quad 
\includegraphics[width=0.46\linewidth]{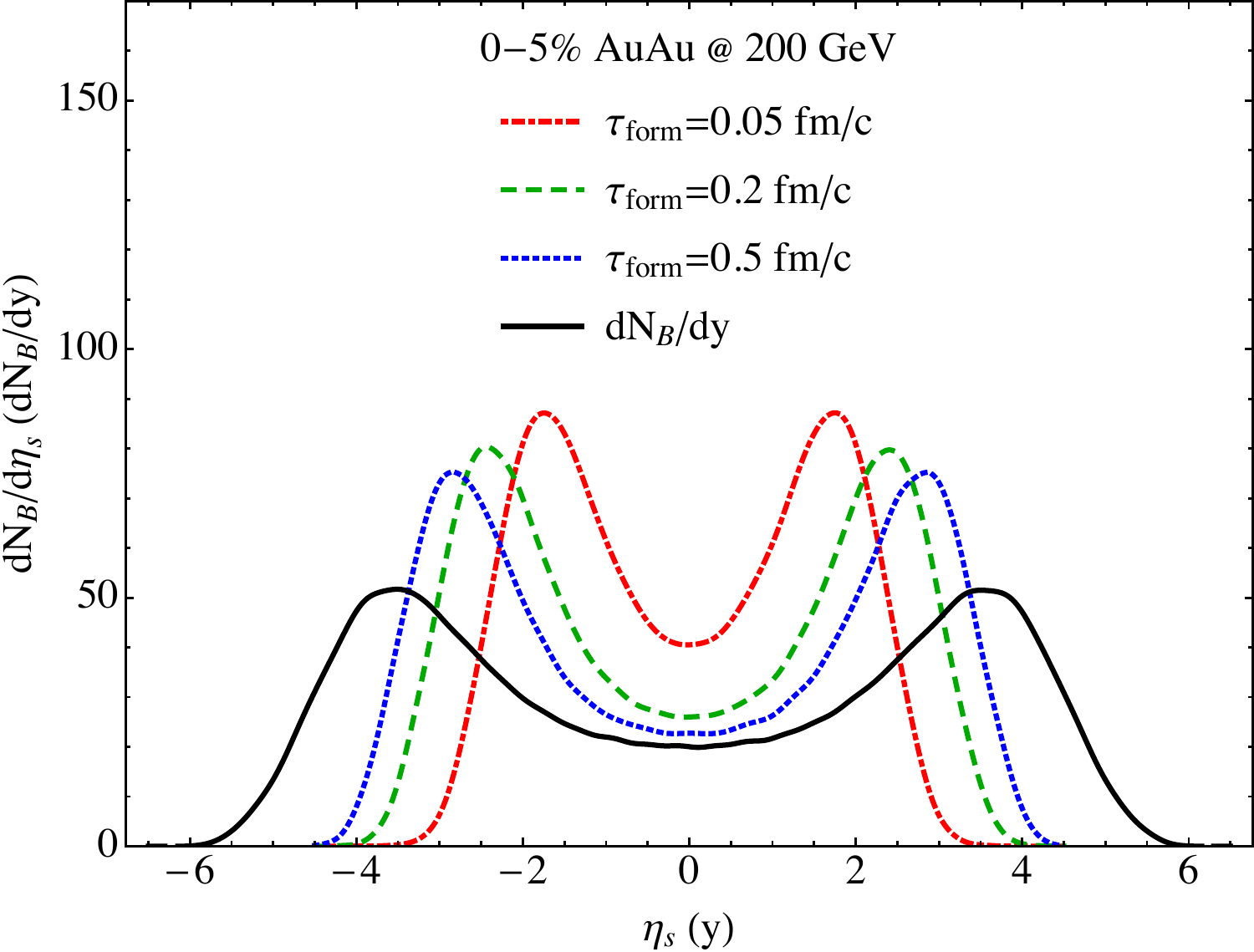}\par 
\caption{{\sl Left:} $\tau{-}\eta_s$ distribution of the produced particles at formation time (mesons not 
	     shown). 
	     {\sl Right:} Space-time rapidity distribution (dashed lines) and rapidity distribution (solid line) of 
	     net baryon number from the pre-equilibrium stage, for different formation times.}
\label{F1}
\end{figure}%
%%%%%%%%%%%%%%%%%%%%%%%%%%%%%%%%%%%%%%%%%%%%%%%%%%
%
In the first expression $\gamma_i=1/\sqrt{1{-}\bp_i^2/(p_i^0)^2}$ and $\bu_i = \bp_i/m_i$ are the Lorentz-contraction factor and spatial part of the four-velocity of a particle with mass $m_i$ and momentum $\bp_i$ in the lab frame. The temporal switching function describes the disappearance and absorption by the fluid of particle $i$ around time $t_\mathrm{rf}=t_{\mathrm{form},i}=t_{i0,\mathrm{rf}}+\tau_\mathrm{form}$ in the particle rest frame; it approaches a step function $\theta(t_{\textrm{form}}{-}t)$ when $\Delta\tau_{\mathrm{th}}\to 0$. We here use a non-zero $\Delta\tau_{\mathrm{th}}{\,=\,}0.5$\,fm/$c$ to avoid large dissipative effects arising from large temporal gradients of the hydrodynamic source terms in Eq.~(\ref{eq1}).

The left panel of Fig.~\ref{F1} shows a scatter plot in the $(\tau,\eta_s)$ plane of the baryons and antibaryons created by UrQMD at their thermalization time $t_{\mathrm{form},i}$, using $\tau_\mathrm{form}=0.2$\,fm/$c$. It shares qualitative features with Fig.~4b in Ref.~\cite{Shen:2017bsr}. In the right panel, the green dashed line converts this information into an initial space-time-rapidity ($\eta_s$) distribution for the hydrodynamic evolution. One observes a large difference between the initial rapidity ($y$, black line) and space-time-rapidity ($\eta_s$) distributions (colored lines), and this difference depends sensitively on the formation time $\tau_\mathrm{form}$. In the dynamical string-fragmentation model of Ref.~\cite{Shen:2017bsr} the initial $y$ and $\eta_s$ distributions for net baryons are much closer to each other, both at $\sqrt{s}=200$ and 19.6\,$A$\,GeV (see Fig.~7 in \cite{Shen:2017bsr}). In our model, at 19.6\,$A$\, GeV the two peaks near projectile and target rapidities merge into a single peak around midrapidity, for both the $y$ and $\eta_s$ distributions, with the width of the $\eta_s$ distribution depending strongly on the choice of $\tau_\mathrm{form}$ but being generically much smaller than that of the rapidity distribution. -- Different initial net-baryon $\eta_s$ distributions correspond to different initial space-time distributions of the baryon chemical potential $\mu_B/T$ whose gradients drive the baryon diffusion current. How the final baryon momentum distributions are affected by the ensuing differences in hydrodynamic evolution is an interesting question. -- We also note that, different from Ref.~\cite{Shen:2017bsr} where the fluctuations in the transverse plane of net baryon and energy densities are correlated with each other and across rapidities by their common string breaking origin, such correlations are not visible in our model, due to the effects of individual transverse and longitudinal motion of the produced particles before becoming part of the fluid.   
 
%%%%%%%%%%%%%%%%%%%%%%%%%%%%%%%%%%%%%%%%%%%%
\section{Dissipative hydrodynamics and baryon evolution}
\label{sec3}
%%%%%%%%%%%%%%%%%%%%%%%%%%%%%%%%%%%%%%%%%%%%

For the dissipative hydrodynamic evolution we solve the dynamically sourced conservation equations 
\begin{eqnarray}
   \partial_{\mu }T^{\mu\nu}_{\textrm{fluid}} = \partial_{\mu } (\ed u^{\mu}u^{\nu} 
   - (\peq+\Pi)\Delta^{\mu\nu}+\pi^{\mu\nu}) = J_{\textrm{source}}^{\nu}\,,
\quad
   \partial_\mu N^{\mu}_{\textrm{fluid}} 
   = \partial_\mu(\n u^{\mu}+\V^{\mu}) = \rho_{\textrm{B}{\textrm{source}}}
\end{eqnarray}
together with relaxation equations \cite{Denicol:2014vaa} for the bulk viscous pressure $\Pi$, shear stress $\pi ^{\mu\nu}$, and baryon diffusion current $\V^{\mu}$. The latter describes a net baryon current in the local rest frame of the momentum flow (Landau frame). Its relaxation equation has the form \cite{Denicol:2014vaa}
\begin{equation}
   \dot{n}^{\left\langle \mu \right\rangle} 
   = -\frac{1}{\tau_n}\left(n^{\mu}{-}\kappa_B \nabla^{\mu}\Bigl(\frac{\mu_B}{T}\Bigr)\right)     
      + \left(\omega^{\mu\nu} -\frac{\lambda _{nn}\sigma^{\mu\nu}+\delta_{nn}\theta g^{\mu\nu}}{\tau_n}\right)n_\nu\;,
\end{equation}
which clearly identifies the gradient of $\mu_B/T$ as the driving force for the baryon diffusion current. 
%
%%%%%%%%%%%%%%%%%%% Fig. 2 %%%%%%%%%%%%%%%%%%%%%%
\begin{figure}[b!]
\centering\includegraphics[width=0.9\linewidth]{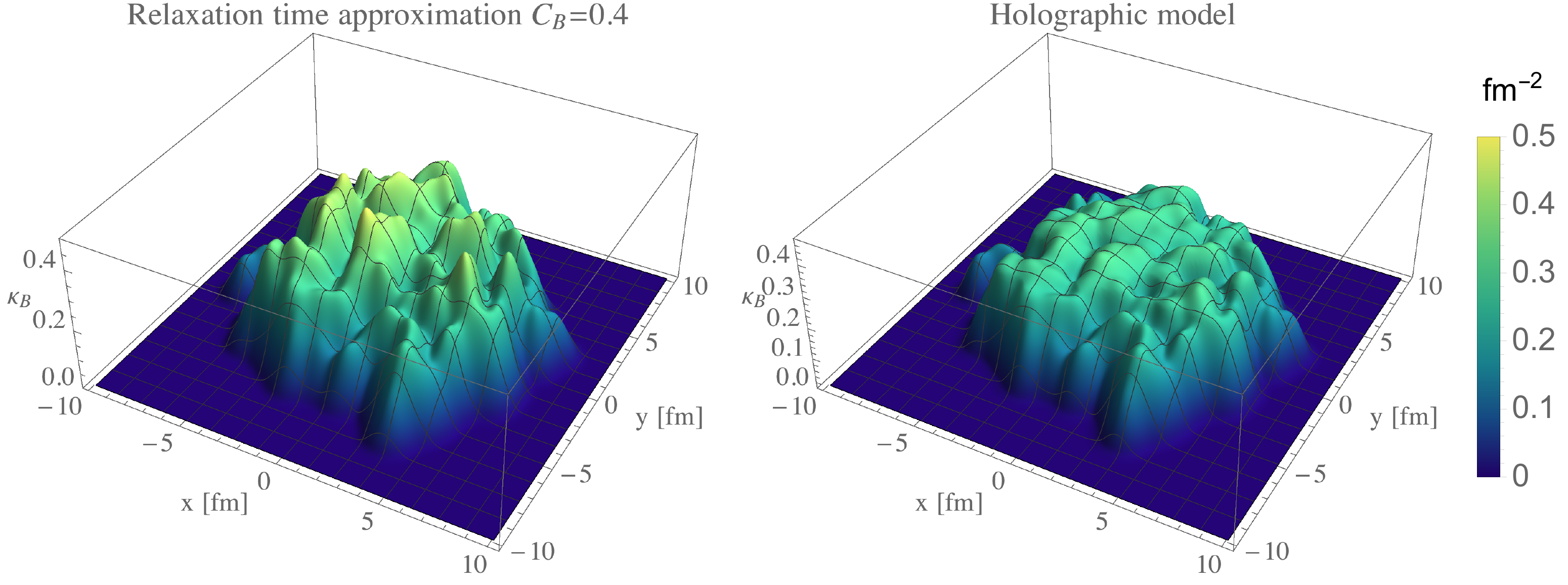}
\caption{Distribution of the baryon diffusion coefficient $\kappa_B$ from kinetic theory in the relaxation time approximation \cite{Denicol:2018wdp} (left panel) and from the holographic model \cite{Rougemont:2015ona} (right panel), for identical MC-Glauber initial energy and baryon density profiles. $C_B=0.4$ lies in the lower half of the typically explored range  \cite{Denicol:2018wdp}.}
\label{F2}
\end{figure}
%%%%%%%%%%%%%%%%%%%%%%%%%%%%%%%%%%%%%%%%%%%%
%
In the Navier-Stokes limit, $n^{\mu}_\mathrm{NS} = \kappa_B \nabla^{\mu}\left(\frac{\mu_B}{T}\right)$, the baryon diffusion coefficient $\kappa_B$ controls the diffusion current created in response to this driving force. Its value depends on the microscopic properties of the medium and has been calculated for weakly coupled massless particles from kinetic theory \cite{Denicol:2018wdp},
\begin{equation}
    \kappa_B = \frac{C_B}{T}\n\left(\frac{1}{3}\coth\left(\frac{\mu_B}{T}\right)-\frac{\n T}{e+\peq} \right),
\end{equation}
and for a strongly coupled medium using a holographic model \cite{Rougemont:2015ona}. Fig.~\ref{F2} compares the initial distribution of $\kappa_B$ in the transverse plane at midrapidity computed from the initial energy and baryon density profiles with MC-Glauber input. Clearly visible significant differences between the two models, in both magnitude and ``bumpiness'', are expected to affect the diffusion of net baryon number and the final baryon spectra.

%
%%%%%%%%%%%%%%%%%%%%% Fig 3 %%%%%%%%%%%%%%%%%%%%%%%%%%%%%%
\begin{figure}[h]
\centering
\includegraphics[width=0.44\linewidth]{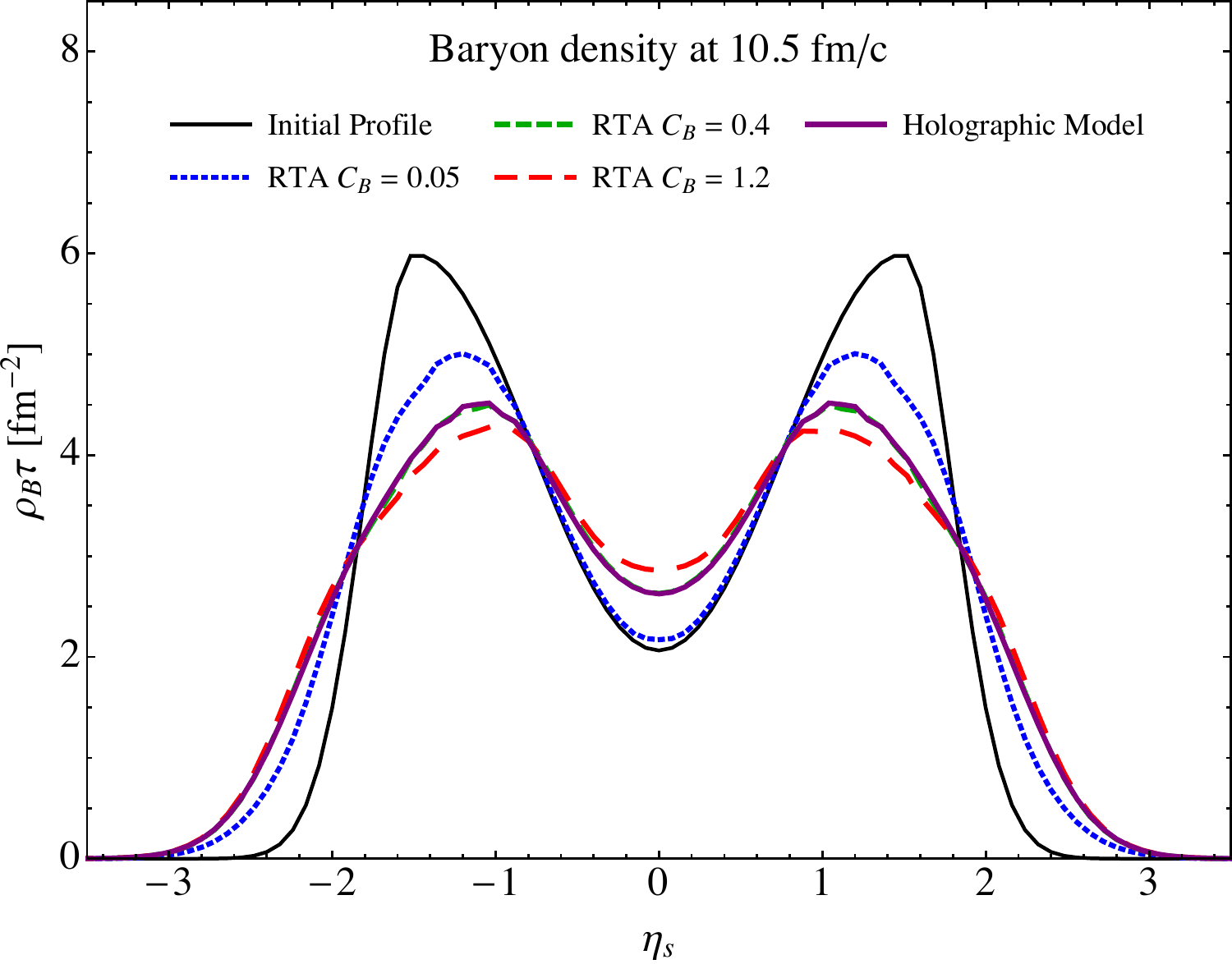}\quad
\includegraphics[width=0.453\linewidth]{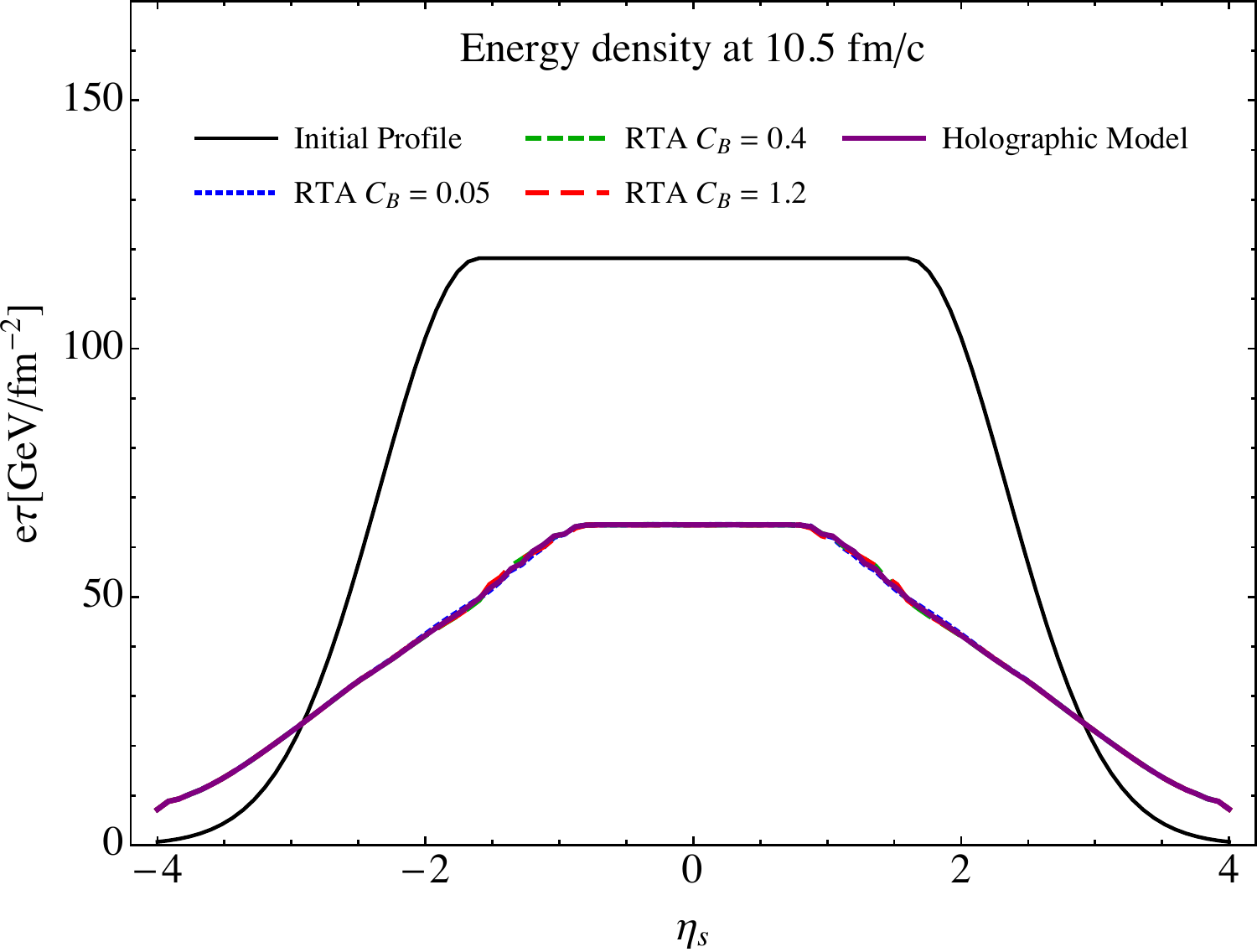}
\caption{For simple initial energy and baryon profiles \cite{Denicol:2018wdp} (black solid lines), the net baryon (left) and energy (right) density distributions in space-time rapidity, multiplied by a factor $\tau$ to counter the dilution effect arising from the initially linear growth in volume associated with 1-dimensional Bjorken expansion along the beam direction, are shown at $t{\,=\,}10.5$\,fm/$c$, using different models for the baryon diffusion coefficient $\kappa_B$ while fixing all other transport coefficients. The evolved $\eta_s$-distributions for the $C_B{\,=\,}0.4$ kinetic theory and holographic models for $\kappa_B$ agree almost perfectly.}
\label{F3}
\end{figure}
%%%%%%%%%%%%%%%%%%%%%%%%%%%%%%%%%%%%%%%%%%%%%%%%%%%%%%
%

To test our (3+1)-dimensional dissipative hydrodynamic code with baryon diffusion we initialized it without dynamical sources (i.e. at constant proper time) using simple initial energy and baryon density profiles \cite{Denicol:2018wdp} (black solid lines in Fig.~\ref{F3}) and zero initial baryon diffusion current. The $\mu_B/T$ gradients associated with the double-humped initial net baryon distribution drive a baryon diffusion current that slightly broadens the initial net baryon space-time rapidity distribution but mostly fills in the initial depression near midrapidity (left panel of Fig.~\ref{F3}), in agreement with Refs.~\cite{Shen:2017ruz, Denicol:2018wdp}. While $\tau$ times the net baryon rapidity density at midrapidity increases with time due to the effect of baryon diffusion which depends strongly on $\kappa_B$, $\tau$ times the thermal energy density at midrapidity decreases with time due to work done by the pressure (right panel of Fig.~\ref{F3}). Due to the small fraction of the total energy carried by the baryons, the energy density evolution is completely insensitive to baryon diffusion.

Results from ongoing studies using the same (3+1)-dimensional dissipative hydrodynamic code with dynamical sources at BES energies will be reported elsewhere.

%%%%%%%%%%%%%%%%%%%%%%%%%%%%%%%%%%%%%%%%%%%%%%%%%%%%%%%%%%%
%% References with BibTeX database:
%%%%%%%%%%%%%%%%%%%%%%%%%%%%%%%%%%%%%%%%%%%%%%%%%%%%%%%%%%%

\bibliographystyle{elsarticle-num}
\bibliography{qm18-du}
\end{document}